\documentclass[10pt,twocolumn]{aastex631}

\usepackage[version=4]{mhchem}
\usepackage[caption=false]{subfig}
\graphicspath{{./}{Figures/}}

\begin{document}

\title{A systematic study of hot O production and escape from Martian atmosphere in response to enhanced EUV Irradiance from Solar Flares}

\author[0009-0004-7149-6442]{Chirag Rathi}
\affiliation{Ohio University \\
Athens, OH, USA 45701}
\altaffiliation{New York University \\
Abu Dhabi, UAE}
\author[0000-0002-9425-2123]{Dimitra Atri}
\affiliation{New York University \\
Abu Dhabi, UAE}

\author[0000-0001-7927-2727]{Dattaraj B. Dhuri}
\affiliation{New York University \\
Abu Dhabi, UAE}









\begin{abstract}


The study of the evolution of Martian atmosphere and its response to EUV irradiation is an extremely important topic in planetary science. One of the dominant effects of atmospheric losses is the photochemical escape of atomic oxygen from Mars. Increasing the magnitude of the irradiation changes the response of the atmosphere. The purpose of the current paper is to analyze the effects of enhanced EUV irradiation on the escape rates of oxygen atoms. We have used the solar flare of $2017$ September $10$ as the baseline flare intensity and varied the intensity of the flare from a factor of $3$ up to $10$ times the baseline flare. We see an increase in the escape flux by $\sim 40 \%$ for flares up to $5 \times$ the intensity of the baseline flare. However, beyond this point, the increase in escape flux tapers off, reaching only about 17\% above the baseline. At $10 \times$ the baseline flare intensity, the escape flux decreases by nearly 25\% compared to the original flare’s escape rate. We also found that the total escape amount of hot $O$ peaks at $7 \times$ the original flare intensity. Additionally, we have studied the effects of the time scales over which the flare energy is delivered. We find that energy dissipative processes like radiative cooling and thermal collisions do not come into play instantaneously. The escape flux from higher intensity flares dominate initially, but as time progresses, energy dissipative processes have a significant effect on the escape rate.

\end{abstract}



\section{Introduction \label{sec:intro}}

Solar flares are sudden brightening in the solar corona that release tremendous amounts of energy (up to $10^{32}$ ergs) in the form of intense short-wavelength radiation, occasionally accompanied by solar energetic particles (SEPs) and coronal mass ejections (CMEs) \citep{Shibata2011,Janvier_Coronal_obs_MHD_sims}. The underlying cause of solar flares is the release of free magnetic energy stored within coronal magnetic fields, which are anchored in active regions on the Sun's surface. This energy is liberated through a process known as magnetic reconnection, which results in the rapid restructuring of these magnetic field lines \citep{parker_cosmo_mag_fields, Su2013}. Eruptive solar flares are often accompanied by CMEs that eject vast amounts of plasma mass into inter-planetary space. When these radiation and plasma are deposited in the planetary atmospheres, such as Earth's and Mars's, they profoundly impact the planetary atmospheres. \\
While the atmosphere of Earth is protected by a strong global magnetic field, the magnetic field of the CME can still cause strong geomagnetically induced currents (GICs) that are known to severely disrupt electrical power grids. In contrast, Mars does not have the protection of a strong global magnetic field, making its atmosphere largely exposed to incoming radiation and plasma from CMEs. However, recent studies like \cite{Sakata_2024} suggest that role of a global magnetic field is more nuanced and may not always prevent atmospheric loss. The interaction of the solar flare with the atmosphere can lead to substantial alterations in its atmospheric composition and dynamics, with potential implications for the planet's long-term atmospheric evolution. Moreover, the flares are ubiquitous to Sun-like magnetically active stars, and can impact the atmospheres of terrestrial planets that can harbor life in an identical manner.
Therefore, understanding and quantifying the effects of radiation on the planetary atmospheres is important.

While solar flares cause increase in the intensity of the radiation across the entire electromagnetic spectrum, their effects are particularly pronounced in the short-wavelength X-ray and EUV regions in which they are routinely observed \citep{Su2013}. The magnitude of the solar flares are therefore classified in terms soft X-ray flux measured using the Geostationary Orbiting Environmental Satellites (GOES) near Earth \citep{Irradiance_obs_goes}. This classification ranks the flares into A, B, C, M and X classes, with the major flares M- and X-classes producing an X-ray flux greater than ${\rm 10^{-5}~W-m^{-2}}$ and ${\rm 10^{-4}~W-m^{-2}}$ respectively \citep{Veronig_temp_aspects_X_ray, Janvier_Coronal_obs_MHD_sims}. 

NASA's Mars Atmosphere and Volatile EvolutioN (MAVEN) mission \citep{Jakosky2015} has been operational since 2014 and one of its goals is to observe the response of the Martian atmosphere to the radiation and plasma from the Sun, particularly during severe space weather. On $2017$ September $10$, a powerful X8.2 solar flare, the second largest during the solar cycle \#24, erupted and heavily impacted Mars. The flare started around $16$:$00$ UT, reached the maximum ionizing irradiance shortly after, and reduced to $50 \%$ of its maximum by $18$:$00$ UT (\cite{Mars_response_10Sept_flare_Theimann}, Figure 1). The events began with two large M-class flares, followed by an X-class flare \citep{EUV_obs_solar_flares_Chamberlin}. This flare caused a dramatic enhancement over a wide range of wavelength including X-ray and EUV. It was initially observed on Earth by GOES X-ray flux instrument and later observed by the Extreme Ultraviolet Monitor (EUVM) on-board in MAVEN \citep{solar_euvm_Eparvier}. This solar flare was one of the largest to occur during the MAVEN mission and a few larger flares have recently occurred during the solar cycle \#25 \cite{Kruparova_2024}. Numerous studies have focused on the effects of the $2017$ flare on the Martian atmosphere \citep{Obs_impact_flare_Lee_2018, Rapid_heating_Mars_Elrod_2018, Mars_response_10Sept_flare_Theimann, MGITM_sims_10Sept_Pawlowski_2019, Effects_of_flare_Cramer_2020, Effects_flare_density_comp_ionosphere_Cramer_2023}. 
Solar flares like the one on $2017$ September $10$, present a rare opportunity
to directly study the changes in the atmosphere composition as a result of enhanced solar ionization and quantify the response of Martian atmosphere in terms of oxygen loss to such extreme space weather events. 
Although other neutrals like $C$, $N$ and $H$ also escape Mars, $O$ is the dominant escaping atom by almost $2$ orders of magnitude (\cite{solar_wind_interaction_book}, Table 3). Accurate models of $O$ escape due to such extreme solar flares are therefore important for understanding the evolution of the planet's atmosphere. 

On Mars, photochemical escape is thought to be the primary loss process of neutrals heavier than hydrogen \citep{Atm_escape_Lammer}. In this escape mechanism, an exothermic reaction produces a neutral particle whose velocity exceeds the planetary escape velocity and the particle is not prevented from escape by subsequent collisions with thermal neutrals. While the photochemical escape of neutrals like $C$, $N$, and $H$ are the result of processes like photodissociation, photodissociative ionization and electron-impact dissociative ionization, dissociative recombination (DR) of $O_2^+$ ions is thought to be the dominant mode of atmospheric loss on Mars for oxygen \citep{photochem_escape_first_results_Lillis, Jakosky_2018_Mars_loss}. In this study too, we consider DR to be the dominant mode of escape for O atoms and our goal is to investigate the response of the Martian atmosphere, in context of $O$ escape, with increasing flare intensities. DR is a process by which hot O atoms are produced in the ionosphere through either of the following channels:

\begin{equation} 
\label{DR_branches}
\begin{split}
    \ce{O_2^+ + e^ &-> O(^3P) + O(^3P) + 6.96 eV \\
     &-> O(^3P) + O(^1D) + 5.00 eV \\
     &-> O(^1D) + O(^1D) + 3.02 eV \\
     &-> O(^1D) + O(^1S) + 0.80 eV.}
\end{split} 
\end{equation}

The critical altitude region for photochemical escape is between $170-250$ km because in this altitude range, substantially large amounts of $O_2^+$ ions are produced and the mean free path is large enough to allow the escape of $O$ atoms. The escape energy of $O$ near an altitude of $200 km$ is $E_{esc} \sim 2$ eV. Each channel of reaction in equation (\ref{DR_branches}) will produce two oxygen atoms and the energy produced will be shared between them. Therefore, the first two channels in equation (\ref{DR_branches}) can produce hot O atoms capable of escaping the Martian atmosphere. Different numerical models like global circulation models \citep{global_circ_model_Bougher_1999, global_circ_model_Bougher_2000, global_circ_model_Gonzalez_2013}, 3D Monte Carlo hot atom transport model \citep{photochem_escape_first_results_Lillis}, simple scaling methods \citep{hot_oxygen_simple_scaling_cravens} etc. have been developed over the years to study the photochemical escape of $O$ from Mars. In this work, we have used the outline of \cite{hot_oxygen_simple_scaling_cravens} to estimate the escape flux of hot $O$ from the planet's atmosphere. This particular method has been chosen due its ability to provide reliable estimates of $O$ escape, given its computational simplicity.

In this paper we present the effects of increased flare intensities on hot $O$ escape from Mars's atmosphere. Section \ref{sec:2} discusses details about the simulations we ran and describes the process of estimating the escape flux from the DR process. Section \ref{sec:3} discusses the effects varying the duration of the super-flares on the $O$ escape rates. We conclude with a discussion of our results in Section \ref{sec:4}.



\section{Method to Estimate the Escape Flux of Oxygen} \label{sec:2}
\begin{figure}[htbp]
    \centering
    \includegraphics[width=\linewidth]{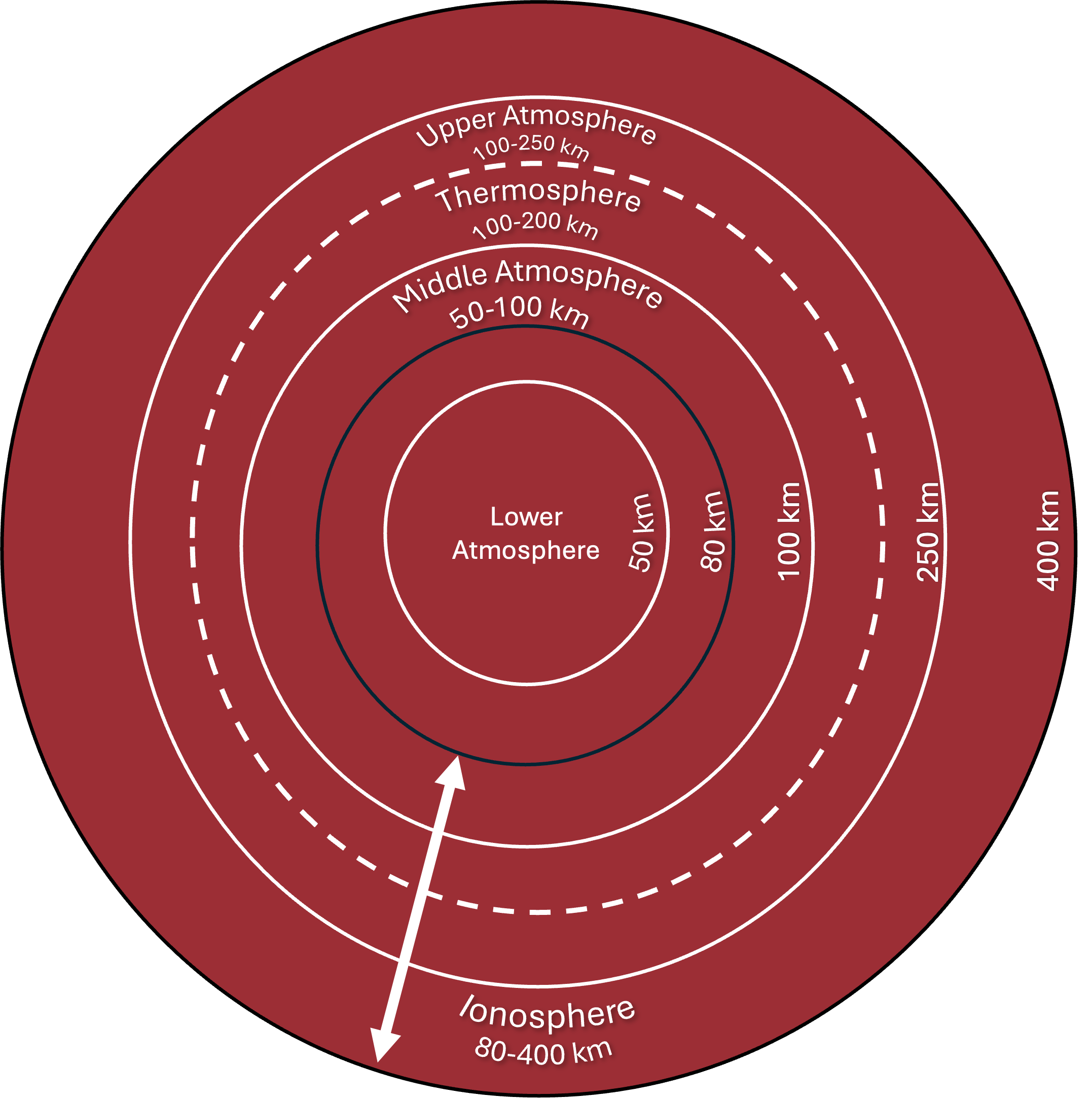}
    \caption{Pictorial representation of the atmospheric layers of Mars. This purpose of this figure is to show the integrated nature of the planetary atmosphere and re-emphasizes the need to study these coupling processes.}
    \label{fig:Mars_layers}
\end{figure}

The Martian atmosphere is generally divided into three segments -- the lower atmosphere ($0-50$ km), middle atmosphere ($50-100$ km) and upper atmosphere ($100-250$ km). 
The Martian ionosphere ($80-400$ km) is largely contained within the thermosphere ($\sim 100-200$ km) and exosphere. The ionosphere is composed of ions like $O_2^+, CO_2^+, O^+, CO^+,$ and $NO^+$ that are formed by the ionization of the neutral thermospheric species. Figure \ref{fig:Mars_layers} shows the integrated nature of Mars's atmosphere through a pictorial representation of the atmosphere's layers. Since the thermosphere-ionosphere (TI) system is strongly coupled to the lower and upper atmosphere, it is affected by the lower atmosphere through dust storms, planetary waves and tides etc. and from the upper atmosphere via solar radiation and solar flares \citep{MGITM_sims_Bougher_2015}. These coupling processes are extremely important to study because they help us predict the behavior of the atmosphere in response to flares. To capture the physics of the entire atmosphere, either one can couple two independent lower atmosphere and upper atmosphere modeling codes or one can employ an integrated framework for the modeling. The Global Ionosphere-Thermosphere Model (GITM) is an integrated three dimensional atmospheric modeling developed to simulate the terrestrial TI system \citep{MGITM_sims_Bougher_2015}. The Mars GITM (MGITM or M-GITM) has been adopted to use for Martian conditions. The MGITM framework combines the terrestrial GITM framework with fundamental Mars physical parameters, ion chemistry and some key radiative processes that capture the important thermal, compositional and dynamical structures in the atmosphere. The list of electron recombination reactions used in MGITM are given below, \\[0.5cm]
\vspace{-1cm}

\begin{align}
    \ce{N2^+ + e^- &-> 2(0.25N(^4S) + 0.75N(^2D))}, \nonumber \\
    \ce{O2^+ + e^- &-> 2O}, \nonumber \\
    \ce{CO2^+ + e^- &-> CO + O}, \\
    \ce{NO^+ + e^- &-> O + N(^2D)}, \nonumber \\
    \ce{NO^+ + e^- &-> O + N(^4S)}. \nonumber
\end{align}


The full list of reactions can be found in \cite{MGITM_sims_Bougher_2015}.

\nolinenumbers

The MGITM model is based on solving the Navier-Stokes equations, which was also implemented in the GITM framework. The model solves these equations in 3D spherical coordinates. In the radial direction the model assumes that each species has its own momentum and continuity equations. Instead of using the standard pressure-based vertical coordinate system, MGITM utilizes an altitude-based vertical coordinate system. Utilizing this approach relaxes the assumptions of hydrostatic equilibrium which becomes extremely important to study the effects of sudden and extreme conditions like solar flares. In the horizontal direction, the GITM framework assumes that all neutral species share the same bulk temperature and energy while all species (neutral and ionic) share the same zonal and meridonial winds in addition to the same background temperature. 

The conditions in the Mars TI system are not constant and heavily depend on the solar activity. The Flare Irradiance Spectral Model (FISM) is an empirical model to estimate the changes in solar UV irradiance due to the solar cycle, solar rotations and solar flare variations \citep{Chamberlin_FISM_2020}. 
Although FISM irradiances are for a distance of 1 AU, they are adjusted in MGITM to reflect the distance to Mars.

The MGITM is fully parallel and can be decomposed into 2D blocks of longitude and latitude. This allows for the model to have a flexible longitudinal and latitudinal resolutions. In this study, the simulations were performed at a high resolution of $2.5 ^{\circ}$ longitude by $2.5 ^{\circ}$ latitude by $2.5$ km altitude. 
We used the results of the work of \cite{Stat_prop_superflares} to run the simulations of enhanced flares. This work analyzed data from $187$ superflares on $23$ solar-type stars. The relation between the duration, $\tau$, and the energy, $E$, of the superflare was found to be,

\begin{equation} \label{time_energy_relation}
    \tau \propto E^{0.39}.
\end{equation} 

We applied this scaling relation to estimate the durations of the enhanced flares analyzed in Section~\ref{sec:extract_altitude_profiles}. These scaled durations serve as a key input in our simulation framework for estimating atmospheric escape rates.

\subsection{Extracting altitude profiles from M-GITM} \label{sec:extract_altitude_profiles}
In order to get any reasonable and reliable estimates of the Martian atmosphere activity during a solar flare, we need to start the M-GITM runs around $60$ days prior to the onset of the flare \citep{MGITM_sims_10Sept_Pawlowski_2019}. This pre-conditioning run is carried out using the spectrum of $2017$ September $3$ assuming constant solar irradiance. The purpose of this run is to spin up the global dynamics to achieve a pseudo-steady state before the flare. From $2017$ September $3$, realistic irradiance spectral model inputs (at $1$ minute cadence) are used for the following days leading up to the main flare event on $2017$ September $10$.

Before calculating the escape profile of oxygen for different flare intensities using MGITM outputs, we first evaluated the performance of MGITM in modeling altitude profiles during the post-flare phase against MAVEN observations. Since this study specifically investigates the post-flare response of the Martian atmosphere, our primary objective was to assess how well MGITM captures atmospheric behavior during this period. With an orbital period of $4.5$ hours, MAVEN had the opportunity to monitor the period after the flare. So, the altitude profiles of its periapsis passage time on $10$ September/$17$:$42$ UT (post-flare) were used. 

\begin{figure*}
     \centering
     \subfloat[\label{fig:e_density}]{
         \centering
         \includegraphics[height=8cm,keepaspectratio]{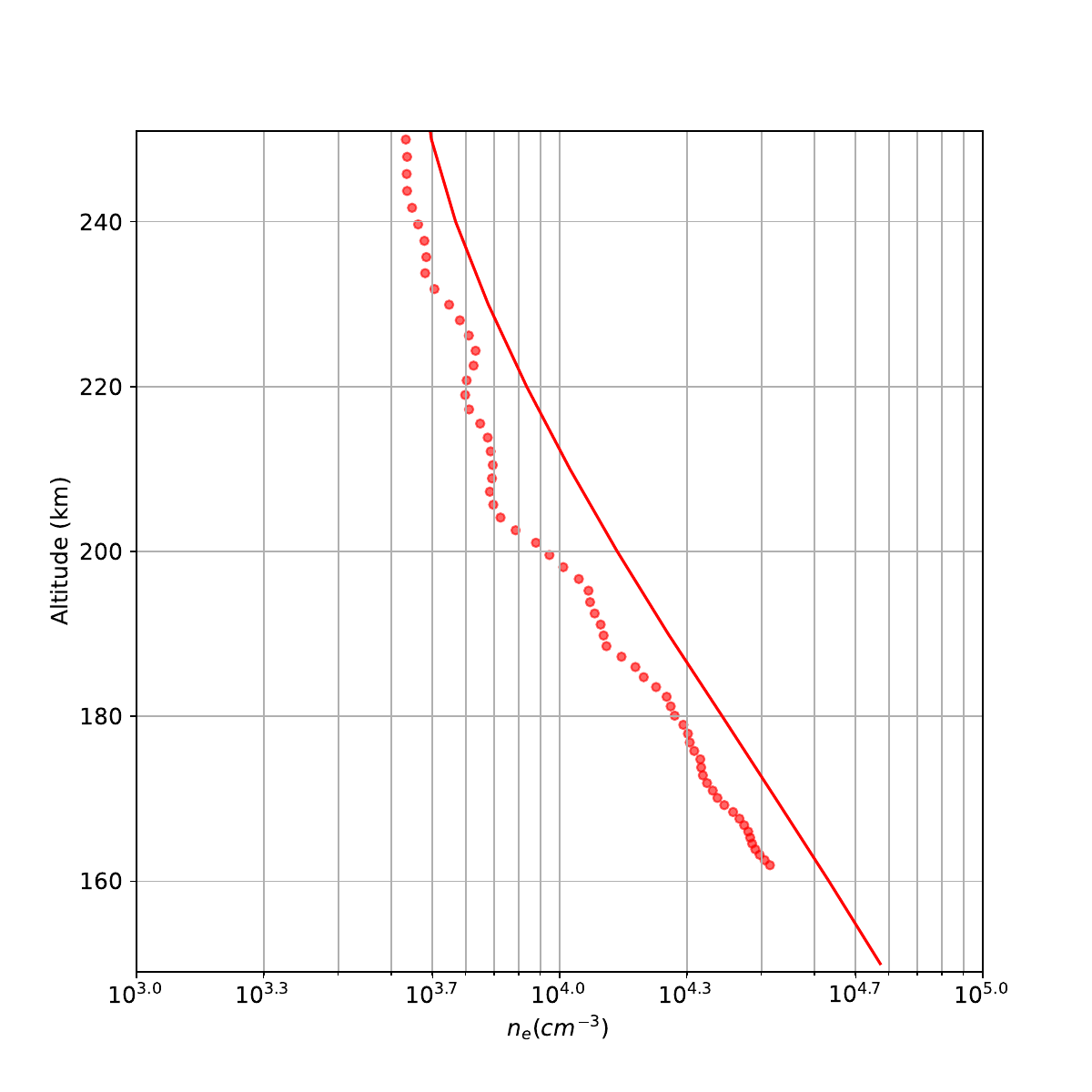}}
     \subfloat[\label{fig:e_temp}]{
         \centering\
         \includegraphics[height=8cm,keepaspectratio]{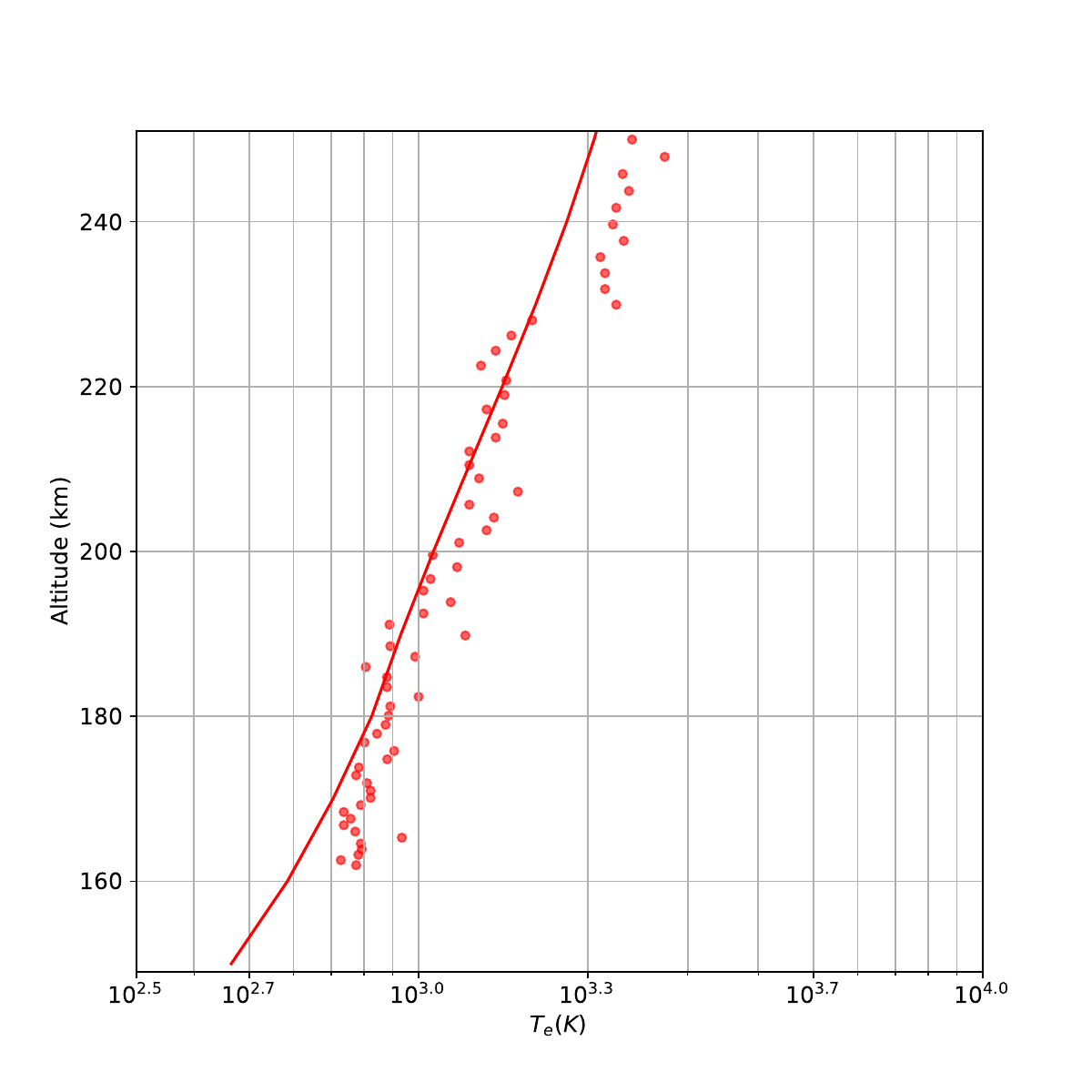}}

     \subfloat[\label{fig:o2_plus_density}]{
         \centering
         \includegraphics[height=8cm,keepaspectratio]{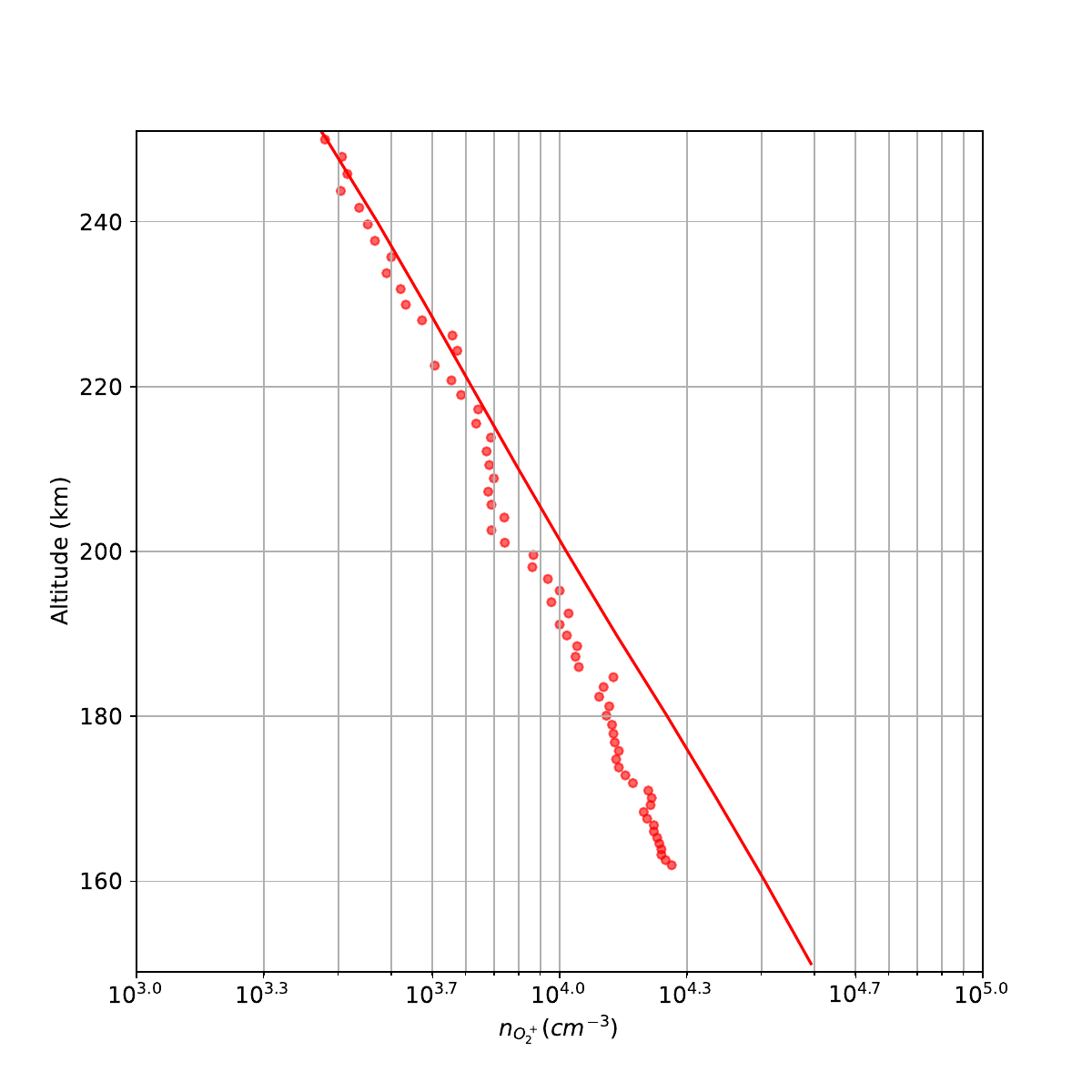}}
         
    \caption{The altitude profiles of three key quantities in calculating the DR rate -- $n_e$, $T_e$ and $n_{O_2^+}$ -- in the post-flare period are plotted. MAVEN observations are plotted in circles while the line plots represent M-GITM simulation results.}
    
    \label{fig:altProfileFigs}
    
\end{figure*}

To evaluate MGITM's fidelity, we focused on three key quantities that govern the DR rate in equation (\ref{DR_rate}): the electron density ($n_e$), the electron temperature ($T_e$) and the $O_2^+$ density ($n_{O_2^+}$). Figure \ref{fig:altProfileFigs} shows the extracted altitude profiles from M-GITM simulations and from MAVEN observations of these key quantities. The electron density and electron temperature were measured using the Langmuir Probe and Waves (LPW) instrument, while the O$_2^+$ density was obtained from the Suprathermal and Thermal Ion Composition (STATIC) instrument aboard MAVEN. The model-data comparison shows that the code generally captures the overall trend of the altitude profiles. However, we notice that there are a few differences between observations and calculations. While the altitude profile of $n_{O_2^+}$ reproduces the observations very well above $\sim 200$ km, the profile generated by MGITM overestimates the $O_2^+$ density at lower altitudes (by $\sim 1.5-2 \times$). The $n_e$ profile shows a systematic difference between the model and observations. Although it captures the overall trend of the profile, it overestimates the density (by $\sim 2 \times$). These discrepancies can be partially attributed to the fact that MGITM, while effective at capturing the thermospheric structure and its response to solar forcing, does not self-consistently solve Maxwell’s equations and therefore lacks a full treatment of electromagnetic fields. These overestimations have implications for the calculation of escape flux. Since the DR rate is proportional to $n_e \cdot n_{\text{O}_2^+}$, the combined effect would result in an overestimation (within an order of magnitude) of the hot O production rate, and therefore the escape flux.


The $T_e$ profile shows remarkable agreement with observations at all altitudes. This agreement is encouraging, although it is important to recognize that the physical processes governing $T_e$ in the ionosphere are inherently complex—especially under disturbed conditions such as those driven by solar flares. In MGITM, electron temperatures are not calculated self-consistently through energy conservation, but are instead prescribed using the empirical formulation by \citet{Fox_1993}, which assumes that the electron temperature equals the neutral temperature below 130~km and increases exponentially above that altitude.

\nolinenumbers

The Martian atmosphere itself is highly dynamic and does not remain in a steady state, even during periods of relatively low solar activity. In the case of the 2017 September 10 event, the flare was closely followed by a fast and wide coronal mass ejection (CME), which was observed shortly after the flare \citep{Lee_2017_flare_CME}. The interaction of the CME with the upper atmosphere could have temporarily altered the electron density through compression or depletion effects in the ionosphere.

Additionally, since our simulations are driven by empirical inputs, the results can be sensitive to the chosen FISM model and the version of the MGITM code. These combined effects—including simplified treatment of $T_e$ and the interaction between plasma and electromagnetic field,  CME-induced variability, and model dependencies—may all contribute to the observed differences in Figure~\ref{fig:e_density}.

         
    
    

\subsection{Estimating the escape flux from DR process}


Studying the Martian atmosphere's response to flares can give us clues in to the planet's past and evolution. Since $O$ is the dominant escaping atom on Mars, studying and quantifying its escape will help us also investigate climate change on the planet. In order to determine the photochemical escape fluxes of hot $O$ atoms from their altitude profiles, we need to determine two quantities as a function of altitude, $z$. They are:
\begin{enumerate}
    \item The production rate of hot $O$ atoms from DR of $O_2 ^+$ ion.
    \item The probability that the hot $O$ atom, once produced, will escape the atmosphere.
\end{enumerate}

Each atom in the pair is produced with half the exothermic energy (momentum conservation) and two $O$ atoms are produced per DR reaction (see equation (\ref{DR_branches})). Therefore, the hot O production rate is,

\begin{equation} \label{DR_rate}
    R = 2n_e(z)n_{O_2^+}(z)\alpha (T_e (z)),
\end{equation}
where the DR coefficient rate (averaged over all possible reaction pathways listed in equation (\ref{DR_branches}). It does not resolve individual product channels but reflects their combined contribution weighted by their respective branching ratios), $\alpha = 1.95$x$10^{-7}$ $cm^3/s$ $(300/T_e)^{0.7}$, and $T_e$ is the electron temperature (in K) at a given altitude, $n_e$ is the electron density and $n_{O_2^+}$ is the $O_2^+$ density \citep{hot_oxygen_simple_scaling_cravens}. Of course, hot $O$ atoms can be produced by mechanisms other than DR, but their contributions to the atoms' escape is in the minority. The probability of escape of a hot $O$ atom depends on the column density and composition of the gas above its location. Hot $O$ atoms produced will collide with other neutral species like $O$, $CO_2$, $N_2$ and $CO$. This will prevent the hot $O$ atoms from escaping and it will heat up the atmosphere. The escape probability as a function of altitude and escape angle, $\theta$, is given by,

\begin{equation} \label{esc_prob_equn}
    G_{esc}(z,\theta) = \exp{\big[-\sum_{i} \sigma_{Oi} N_i \sec{(\theta)}\big]}, 
\end{equation} 

where $i$ represents the species $\textrm{O, CO,} \, \textrm{N}_2, \, \textrm{or CO}_2$. So, $\sigma_{Oi}$ are the backscattering elastic cross sections for O collisions with $O$, $CO$, $N_2$ and $CO_2$ respectively \citep{hot_oxygen_simple_scaling_cravens}. $N_i$ is the vertical column density of a given species. The escape angle for this work was chosen following \cite{hot_oxygen_simple_scaling_cravens} to be $60 ^{\circ}$. While we consider only $4$ major species for oxygen collision (because they are the dominant channels of collisions), one could add more species to the equation \citep{hot_oxygen_simple_scaling_cravens}. 

The first key ingredient, the column densities, were obtained by summing over the number densities (or altitude profiles). These altitude profiles were extracted between $150$ and $300$ km (at $10$ km interval) from the MGITM runs. Since MGITM uses the same photochemical and transport schemes for all flare scenarios, the degree of overestimation in $n_e$ and $n_{O_2^+}$ (as seen in Figure~\ref{fig:altProfileFigs}) is expected to remain approximately consistent across different flare intensities.

To evaluate these escape trends across different flare intensities, we simulated a set of enhanced flare scenarios based on the 2017 September 10 event. The baseline flare event of $2017$ September had clearly entered the post-flare phase by $18$:$00$ UT, with irradiance reduced by more than $50\%$ from its peak. We defined the duration of the original flare as $2$ hours (from $16$:$00$ to $18$:$00$ UT) and used this as a reference point. The durations of other enhanced flares were then scaled accordingly using equation~(\ref{time_energy_relation}). So, if the baseline flare of $2017$ lasted for $2$ hours, the enhanced flares, $3 \times, 5\times, 7\times$ and $10 \times$ would have lasted for approximately $3$ hours, $3.75$ hours, $4.24$ hours and $4.9$ hours respectively. To simulate enhanced flare scenarios, we scaled the EUV irradiance time series from the 2017 September 10 event using multiplicative factors of $3\times$, $5\times$, $7\times$, and $10\times$ relative to the baseline flare. The scaling was applied only to the flare enhancement portion of the time series, while the pre-flare irradiance levels were left unchanged.

The second ingredient for this estimation is the oxygen scattering cross sections with various species. Unfortunately, a lot of discrepancies exist when it comes to oxygen scattering cross sections and this results in wide deviations in the estimation of the escape rates. For instance, many modelers have used the theoretical $O-O$ scattering cross sections calculated by \cite{O_cross_section_Kharchenko_2000} and then have scaled these cross sections to other target species. \cite{O_escape_rates_Fox_Hac_2014} studied the escape of $O$ from Mars using different values of $O$ elastic cross sections with different species. 
Following the work of \cite{hot_oxygen_simple_scaling_cravens}, we have used the back-scattering elastic $O$ cross sections with various species and they are tabulated in Table \ref{tab:cross_section_table}. This simple scaling model treats all major collisions in equation (\ref{esc_prob_equn}) on equal footing and assumes that a single collision can make the $O$ unworthy of escape. This estimation also assumes that ionization is locally produced, which in turn produce hot O atoms via the DR process.

\begin{figure*}
    \centering
    
    \subfloat[\label{fig:esc_prob}]{
         \centering
         \includegraphics[width=0.5\linewidth]{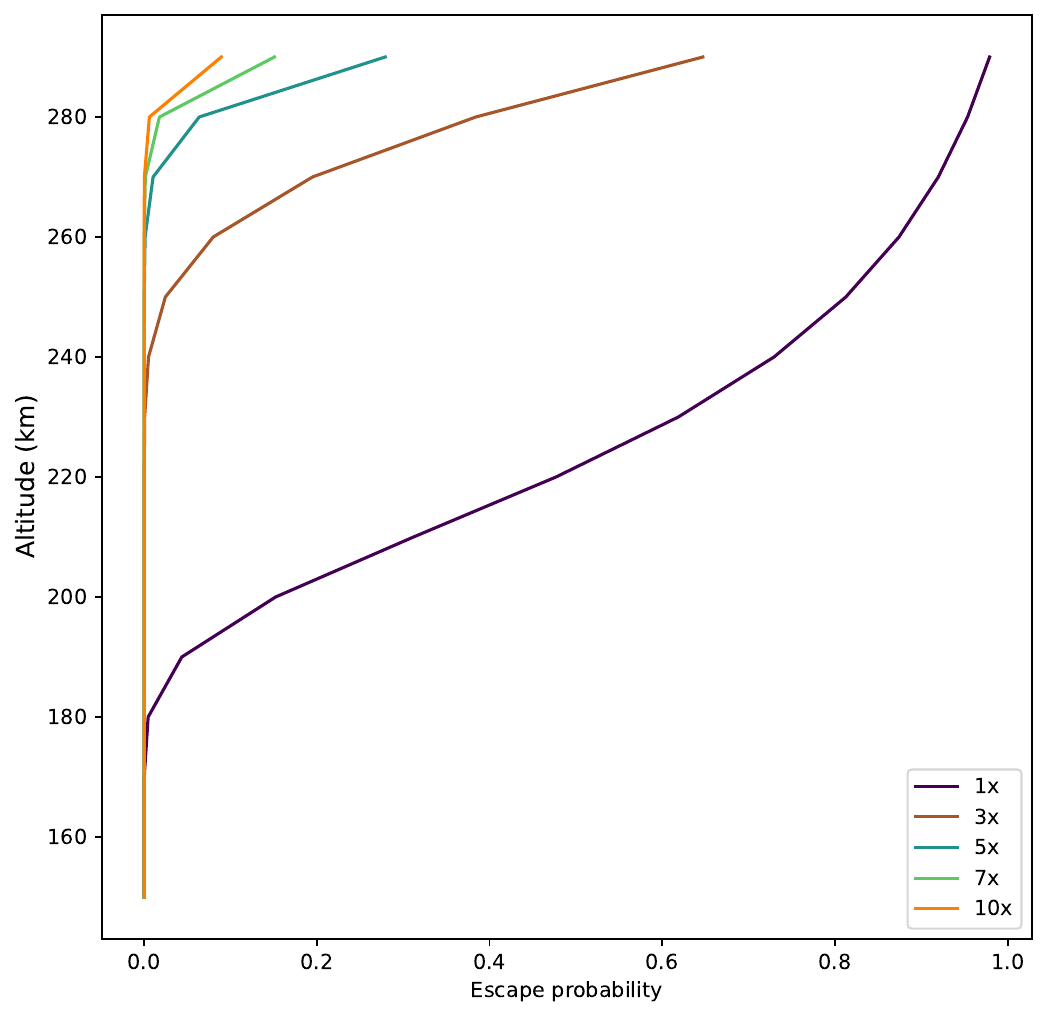}}
    \subfloat[\label{fig:esc_flux}]{
         \centering
         \includegraphics[width=0.5\linewidth]{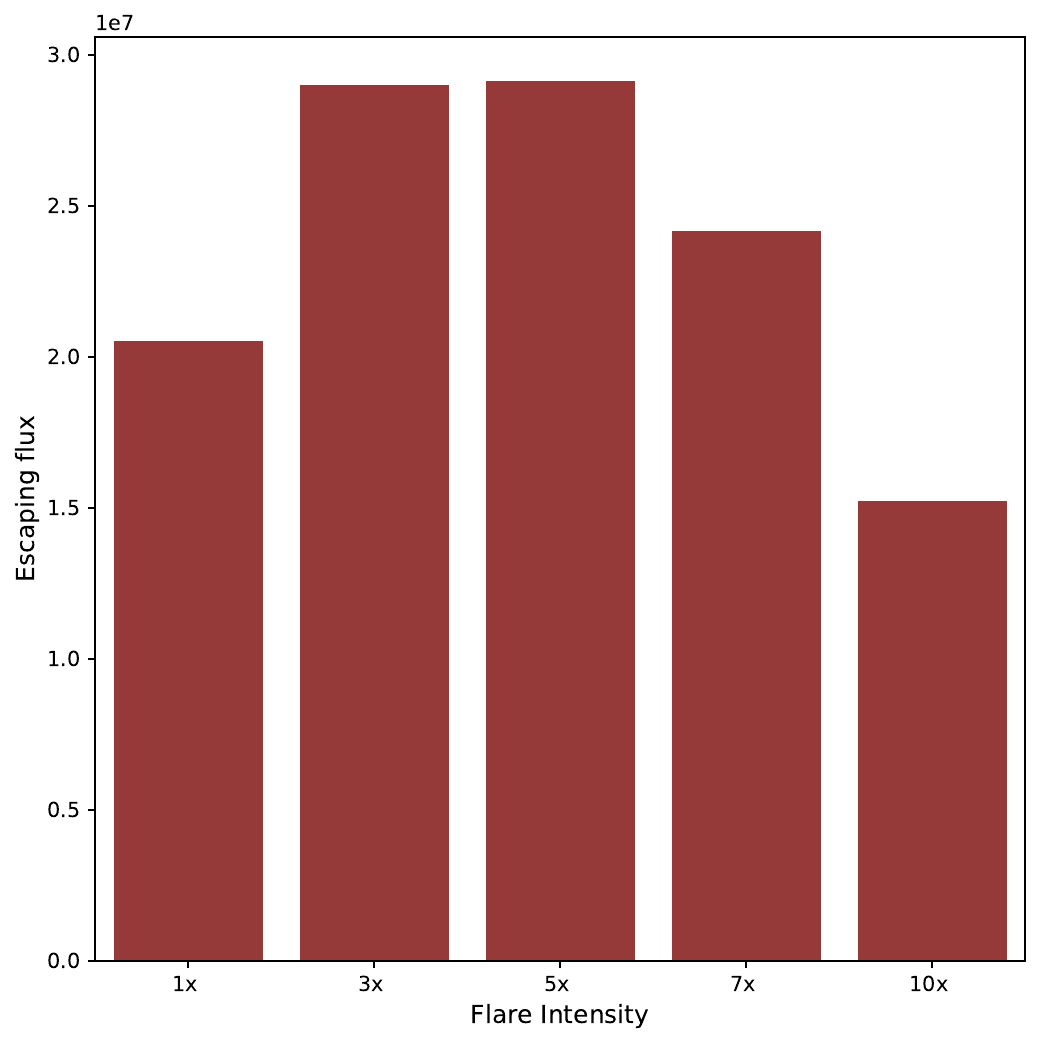}}
         
    \caption{Left: escape probabilities of hot $O$ with increasing flare intensities. As the intensity of the flare increases, it gets harder for the produced hot $O$ atoms to escape. As a result, the probability becomes non-zero only at higher altitudes. Right: The escaping flux (in $cm^{-2} \, s^{-1}$) of hot $O$ at different flare intensities. We see that the flux peaks at $5 \times$ the flare intensity.}
    \label{fig:esc_prob_fig}
\end{figure*}

\begin{figure}
    \centering
    \includegraphics[width=\linewidth, keepaspectratio]{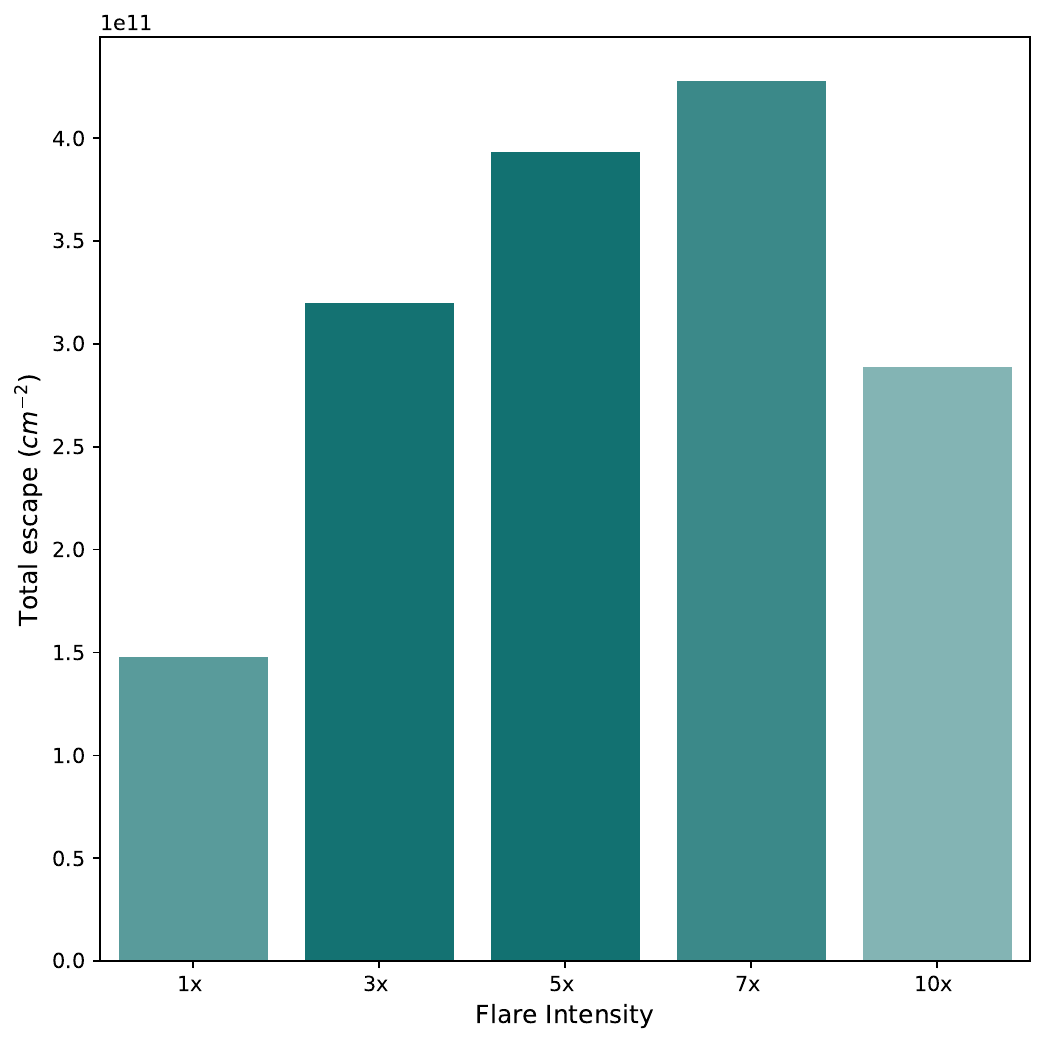}
    \caption{Calculated total escape amount of hot $O$ atoms for different flare intensities. The opacities of the bars in the plot are scaled by the ratio of the escape rate at a given flare intensity to the sum of the escape rates at all intensities. So, a higher escape rate corresponds to a darker color of the bar. It can be seen that although the total escape amount peaks at $7 \times$ flare intensity, the escape rate peaks at $5 \times$ flare intensity. This plot acts as an important tool to visualize the lagging relationship between the total escape amount and the escape rates for a given flare intensity.}
    \label{fig:total_esc_amount}
\end{figure}

Nevertheless, it can provide a very good estimate of the escape rates with increasing flare intensities. Substituting these quantities into equation (\ref{esc_prob_equn}) gives us the escape probabilities for different flare intensities. To calculate the escape production rate, we multiply the hot production rate from equation (\ref{DR_rate}) by the escape probability from equation (\ref{esc_prob_equn}). To obtain the escaping flux, we summed the escape production rate over all altitudes between $150$ and $300$ km. In Figure \ref{fig:esc_prob_fig} we observe that as the intensity of the flare increases, the escape probability shifts to non-zero values at higher altitudes. This would imply that the hot $O$ atoms produced at lower altitudes find it harder to escape the atmosphere with increasing flare intensity. This is because, even though the increased flare intensity leads to enhanced ionization and thus increased production of O$_2^+$ and electrons, it also raises atmospheric density and potentially electron temperature ($T_e$). While higher $T_e$ reduces the dissociative recombination rate coefficient ($\alpha \propto T_e^{-0.7}$), the overall production of hot O atoms can still increase due to elevated $n_e$ and $n_{\text{O}_2^+}$. However, the denser atmosphere simultaneously increases the frequency of collisions, thereby reducing the mean free path and diminishing the escape probabilities of hot O atoms. We estimated the escaping flux to be $2 \times 10^{7}$ $cm^{-2} s^{-1}$, similar to \cite{Mars_response_10Sept_flare_Theimann}, that estimated the flux around $1.7 \times 10^7$ $cm^{-2} s^{-1}$ from the original flare event. As the intensity of the flare increases, we initially see a jump in the escaping flux of $O$ atoms up until $5 \times$ intensity. Beyond that, we see the escape flux drop drastically with increasing intensity because even though we have produced a lot more hot $O$ atoms, the escape probability takes on non-zero values only above $270$ km. However, that is not the case with the total escape amount of hot $O$ atoms. In Figure \ref{fig:total_esc_amount}, the color opacity in the bar plot reflects the escape flux, with darker shades indicating higher escape rates. While the peak escape rate occurs at $5 \times$ the baseline flare intensity, the total escape amount reaches its maximum at $7 \times$ the baseline flare intensity. This lag suggests that the total escape amount is influenced not only by the escape rate but also by the duration of elevated flare intensity. Therefore, both the escape rate and the duration over which the flare intensity is sustained contribute to the total escape amount. 

\nolinenumbers

\begin{table}[htbp]
    \centering
    \begin{tabular}{|c|c|}
        \hline
         Species & Cross section ($cm^2$) \\
         \hline
         $\sigma_{OO}$ & $2.5 \times 10^{-16}$  \\
         $\sigma_{OCO}$ & $9.7 \times 10^{-16}$  \\
         $\sigma_{ON_2}$ & $9.7 \times 10^{-16}$  \\
         $\sigma_{OCO_2}$ & $1.3 \times 10^{-15}$  \\
         \hline
    \end{tabular}
    \caption{Backscattering elastic cross sections of $O$ collisions with other species. These cross sections have been taken from \cite{hot_oxygen_simple_scaling_cravens}.}
    \label{tab:cross_section_table}
\end{table}

\section{Effect of flare duration on hot \textit{O} escape flux} \label{sec:3}
In the previous sections, we analyzed how flare intensity influences the hot \( O \) escape rates. However, intensity alone does not fully determine whether escape rates are enhanced or suppressed compared to the baseline flare. Another crucial factor is the duration over which the flare’s energy is delivered. In this section, we explore the impact of varying flare durations on the escape rates by examining four hypothetical scenarios. The flare durations used in this section are hypothetical and are not drawn from equation (\ref{time_energy_relation}). We wanted to investigate how would the modified duration for each of the enhanced flares affect the escape flux of hot O atoms. Each scenario investigates how the Martian atmosphere responds when the same flare intensity is distributed over different timescales.

To carry out this study, we use enhanced flare intensities of \(3\times\), \(5\times\), \(10\times\), and \(12\times\) and we chose to run simulations for durations of $30, 60, 90$, and $120$ minutes, for each intensity. The goal is to assess how the compression of the flare spectrum in temporal space, affects the hot \( O \) escape flux. 

\begin{figure}
    \centering
    \includegraphics[width=\linewidth]{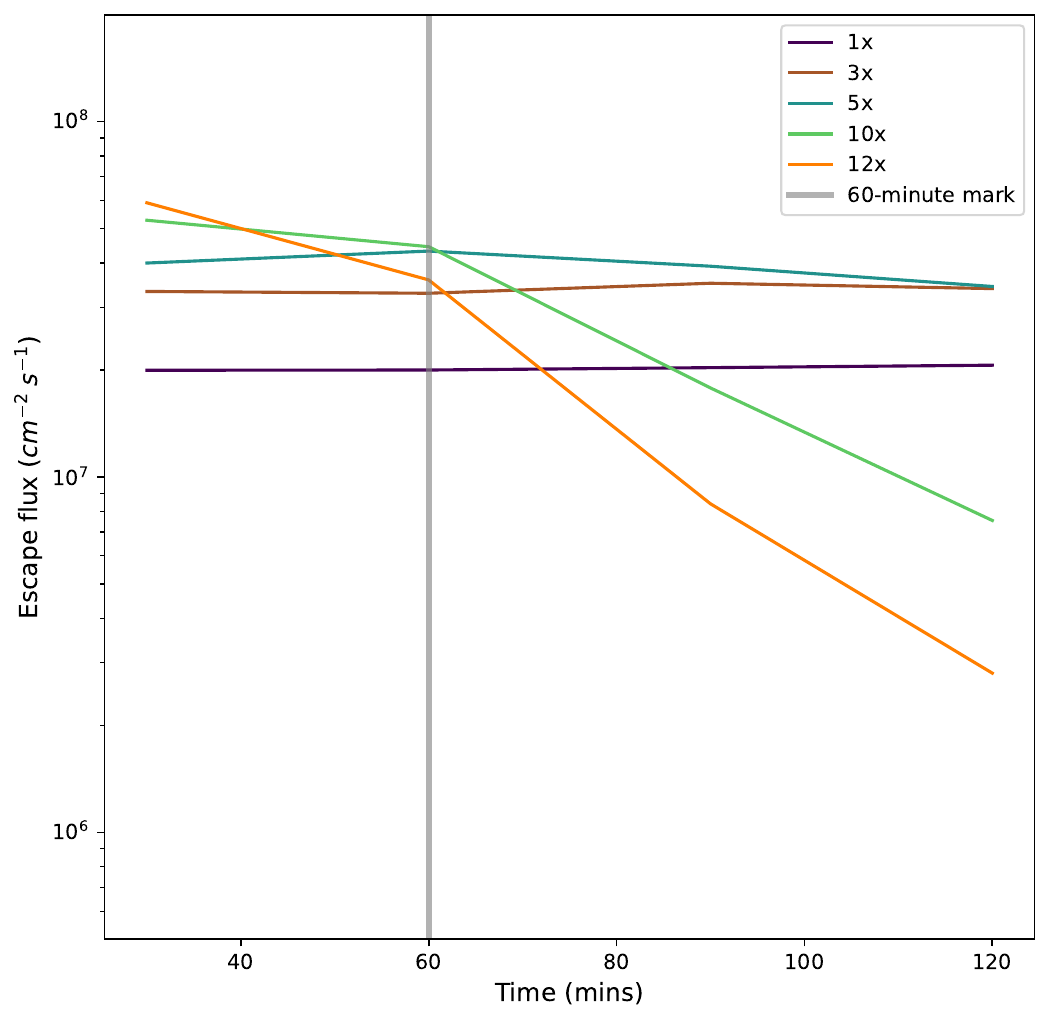}
    \caption{The escape flux as a function of time for different flare intensities.}
    \label{fig:hot_O_escape_time}
\end{figure}

\begin{figure}
    \centering
    \includegraphics[width=\linewidth]{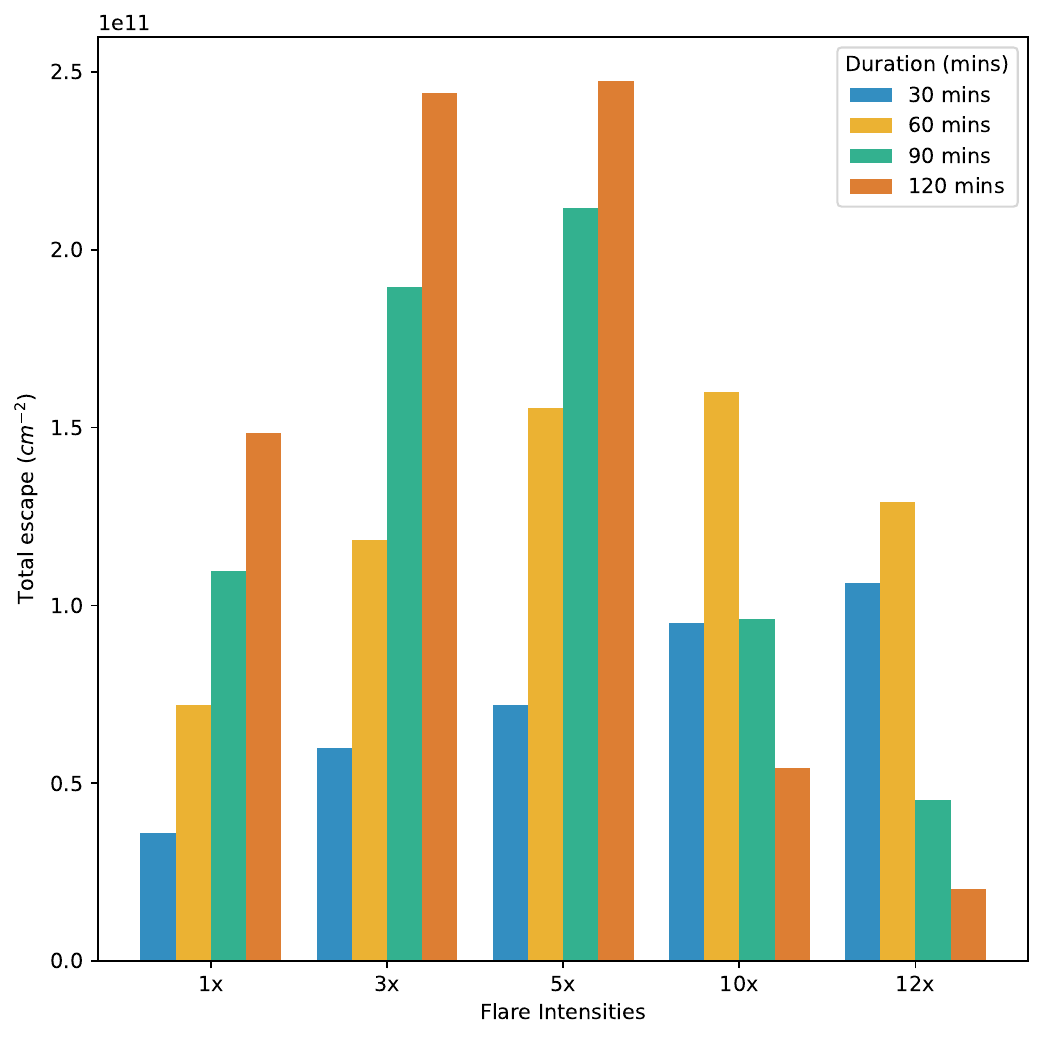}
    \caption{The total hot O escape amount as a function of intensities for different flare durations.}
    \label{fig:hot_O_escape_intensities}
\end{figure}

The purpose of this analysis is to understand the role of flare duration on the hot \( O \) escape process. By compressing the flare spectrum temporally over different timescales, we aim to uncover how the rate of energy delivery influences atmospheric escape. The escaping flux and the total escape amount are presented in Figures \ref{fig:hot_O_escape_time} and \ref{fig:hot_O_escape_intensities}, respectively. We would like to point out two notable features from this analysis.

\begin{enumerate}
    \item \textbf{Initial Peak of High-Intensity Flares}: The escape flux of the $12 \times$ flare initially peaks above all other intensities but declines sharply around 45 minutes, eventually dropping below the $10 \times$ intensity and, by the 60-minute mark, falling beneath the $5 \times$ flare flux. This rapid decline is also apparent in the total escape profile. For short durations, such as the 30-minute period (blue bars in Figure \ref{fig:hot_O_escape_intensities}), the $12 \times$ flare dominates total escape. However, at 60 minutes (yellow bars in Figure \ref{fig:hot_O_escape_intensities}), the total escape amount from the $12 \times$ flare dips below that of the $10 \times$ and $5 \times$ flares. Beyond the 60-minute mark, the total escape rate from the $12 \times$ flare continues to plummet steeply, underscoring a significant decrease in sustained escape potential at high intensities.

    \item \textbf{Shifting Dominance Over Time}: The timing of maximum escape flux and total escape amount varies with flare intensity. For instance, the $12 \times$ flare initially leads in the escape rate and quantity until approximately 45 minutes. Between 45 and 60 minutes, the $10 \times$ flare overtakes it, and after 60 minutes, the $5 \times$ flare becomes the dominant contributor for the remainder of the instances. This pattern indicates a temporal shift in peak escape dynamics: as time progresses, the flare intensities that yield the highest escape rates transition from $12 \times$ to $10 \times$ and eventually stabilize at $5 \times$, marking a sustainable threshold for prolonged escape flux and cumulative escape.

\end{enumerate}

\nolinenumbers

There is always a competition between gaining enough energy to escape and losing energy to various mechanisms like collisions and radiative cooling. These energy loss mechanisms take time to show their effects on the escape of hot $O$ from the atmosphere. A sudden burst of energy imparted on the atmosphere expands the thermosphere and accelerates the particles to higher energies. So, at shorter time scales, we see escape flux and total escape amount from $12 \times$ and $10 \times$ flare intensities dominate. However, as time progresses, the atmosphere contracts to a pseudo-steady state and energy loss mechanisms start to compete with energetic atoms trying to escape. As a result, the flux and total escape amount decline rapidly for $12 \times$ and $10 \times$ flare intensities while for $5 \times$ and $3 \times$, the dip is shallower and takes more time. 

\begin{figure}
    \centering
    \includegraphics[width=\linewidth]{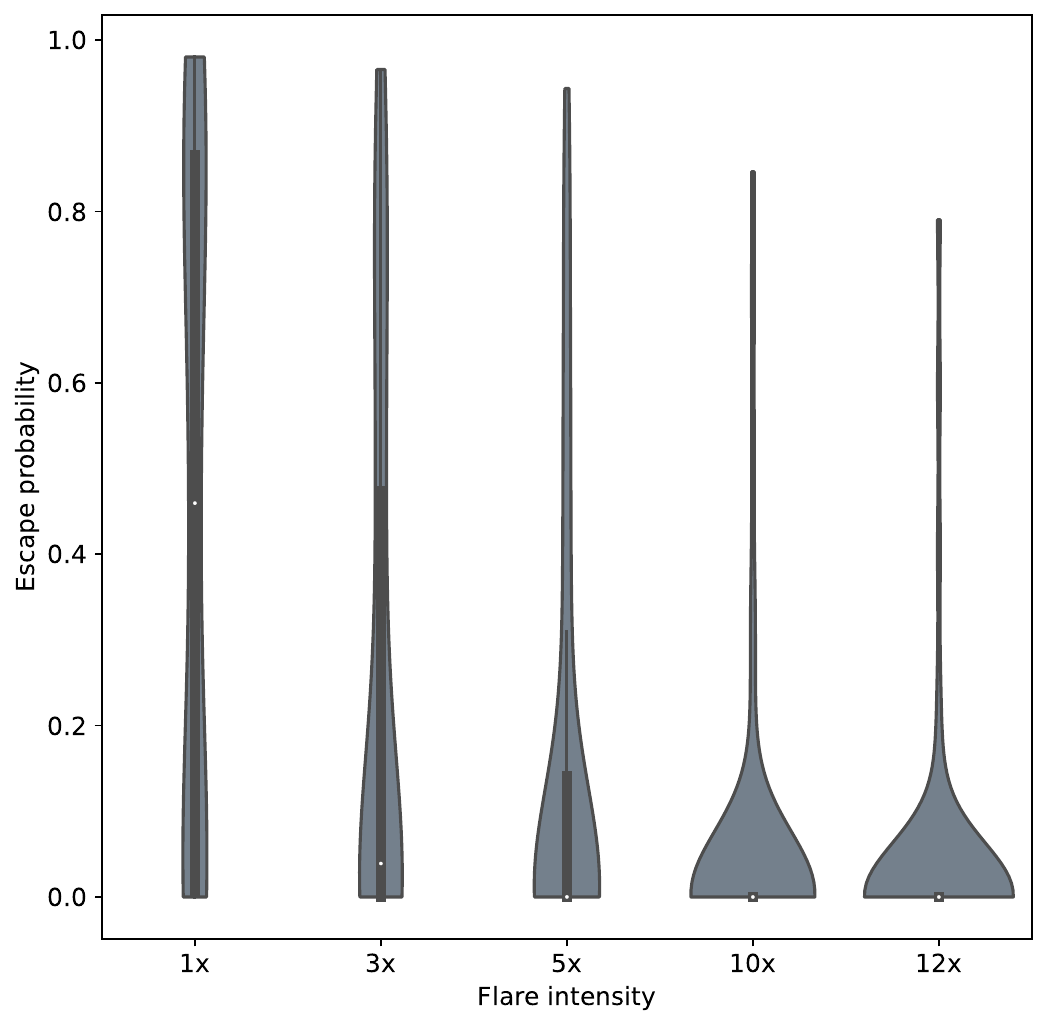}
    \caption{The violin plot is an independent representation of the distribution of calculated escape probabilities across different flare intensities. The distributions illustrate the influence of flare intensity on the probability of atmospheric escape, capturing variations in response to different exposure durations. The width of each violin represents the density of escape probability values, highlighting the spread and central tendency across different conditions. Wider sections of a violin indicate a higher density of hot O atoms with that specific escape probability, meaning more atoms are likely to have that probability of escape.}
    \label{fig:hot_O_escape_prob_fluence}
\end{figure}

Figure~\ref{fig:hot_O_escape_prob_fluence} shows a violin plot generated from the escape probability data, illustrating the statistical distribution of escape probabilities at different flare intensities. The x-axis represents the flare intensities ($1\times, 3\times, 5\times, 10\times, 12\times$), while the y-axis represents the escape probability. Each violin aggregates data from four different flare durations ($30, 60, 90,$ and $120$ minutes) to show how escape probability varies with flare intensity.
The violins depict the probability density distribution of escape probabilities for each flare intensity. The white circle within each violin represents the median escape probability, indicating the value below which $50 \%$ of the computed escape probabilities lie. The dark gray bar inside each violin spans the interquartile range (IQR), covering the middle $50 \%$ of escape probability values (from the first quartile Q1 to the third quartile Q3). The width of the violin at a given escape probability value reflects the frequency of that probability in the dataset, meaning wider regions correspond to escape probabilities that occur more frequently.

In the $1\times$ flare intensity simulation, escape probabilities are relatively evenly distributed around $50 \%$, indicating a moderate likelihood of escape across many cases. However, as the flare intensity increases, the distribution of escape probabilities shifts: the violin plot becomes broader at lower escape probabilities and narrower at higher escape probabilities. This trend signifies that at higher flare intensities, a greater fraction of cases result in low escape probabilities, while fewer cases retain high escape probabilities.
This behavior suggests that while stronger flares inject more energy, they also introduce additional energy loss mechanisms (e.g., collisional thermalization, radiative cooling) that hinder escape. Consequently, at flare intensities of $5\times$ and higher, the median escape probability is near zero, meaning that in at least $50 \%$ of cases, escape is highly suppressed. This trend supports the conclusion that increasing flare intensity does not linearly increase escape flux but instead leads to enhanced trapping of particles due to competing loss processes.

At shorter time scales, the particles initially experience stronger escape fluxes under intense flares, as indicated by the thin body of the violin near higher escape probabilities. However, this effect is transient, as shown by the narrow upper tails. Over time, the escape probabilities shift further downward, concentrating more particles toward lower escape probabilities, which is evident from the broad lower bodies of the violins for higher intensities. As the flare intensity increases, the number of hot $O$ atoms produced also increases. Although the escape rate from stronger flares is higher initially, energy loss mechanisms eventually come into play and pull the escape probability of the atoms to lower values. Thus, as flare intensity increases, the overall escape flux diminishes, with most particles having a less than 50\% probability of escaping. This transition is reflected in both the shrinking upper tails and the broadening lower bodies of the violin plots, especially for the highest flare intensities.


\nolinenumbers

\pagebreak

\section{Discussions \& Conclusions} \label{sec:4}

This study investigates the response of the Martian atmosphere to enhanced solar flare intensities, focusing on the photochemical escape of hot atomic oxygen. By employing the MGITM framework and methodologies from \cite{hot_oxygen_simple_scaling_cravens}, we quantified the escape flux and total escape amount across varying flare intensities and durations. It is important to note that our results are derived within the scope and limitations of the MGITM model; a more comprehensive treatment incorporating the full range of physical processes affecting plasma behavior in the Martian atmosphere may be necessary to accurately capture the atmospheric response to enhanced EUV radiation from solar flares.

Our results demonstrate that while the escape flux of oxygen initially increases with flare intensity, a saturation point occurs beyond 5$\times$ the baseline flare intensity, after which the escape flux declines. This behavior arises from competing processes: while higher flare intensities produce more energetic oxygen atoms, increased atmospheric density at higher altitudes reduces the mean free path for escaping particles due to collisions. Interestingly, the total escape amount of oxygen peaks at 7$\times$ the baseline intensity, indicating that the cumulative escape is influenced by both production rates and flare duration.

Therefore, we also examined the role of flare duration in determining the escape flux and total escape amount. Short-duration, high-intensity flares initially drive significant escape rates due to rapid energy deposition and thermospheric expansion. However, as flare duration increases, energy loss mechanisms such as radiative cooling and collisions counteract the escape process, particularly for extreme intensities (10$\times$ and 12$\times$). This results in a marked decline in escape flux over longer durations. Lower intensities (e.g., 3$\times$ and 5$\times$) exhibit a shallower decline, suggesting a more sustained escape potential.

These findings highlight the complexity of Martian atmospheric dynamics under varying solar activity conditions. The interplay between energy deposition, atmospheric density, and energy dissipation mechanisms governs the escape process, making it highly sensitive to both the intensity and duration of solar flares. Importantly, the results challenge the assumption that larger flares always drive higher escape rates, demonstrating that energy loss mechanisms can significantly offset the escape potential at extreme intensities.

This study contributes to our understanding of the Martian atmosphere's evolution by providing constraints on atmospheric loss processes during extreme space weather events. The insights gained here are not only relevant for Mars but also have implications for other terrestrial planets exposed to high-energy stellar activity, offering a framework to assess atmospheric escape and habitability potential in such environments. 

\begin{center}
    \textbf{Acknowledgments}
\end{center}

This work was supported by the New York University Abu Dhabi (NYUAD) Institute Research Grants CG014, G1502 and the ASPIRE Award for Research Excellence (AARE) Grant S1560 by the Advanced Technology Research Council (ATRC). We would like to thank Dr. Dave Pawlowski at Eastern Michigan University for providing the M-GITM code and for his support while we were learning to use the code. This work utilized the High Performance Computing (HPC) resources of NYUAD. We highly appreciate the support of HPC staff at NYUAD who helped us install and run the code on the Jubail supercomputing cluster.

\begin{center}
    DATA AVAILABILITY
\end{center}
\small
The MAVEN data used for comparison with MGITM in Figure \ref{fig:altProfileFigs} was obtained from the Maven Science Data Center. Specifically, the electron density and temperature were extracted from the LPW instrument and the $O_2^+$ density was extracted from the STATIC instrument. The data is available and can be accessed at \url{https://lasp.colorado.edu/maven/sdc/public/data/sci/kp/insitu/}

\bibliography{sample631}{}
\bibliographystyle{aasjournal}

\end{document}